\documentclass[reprint]{revtex4-2}
\usepackage[T1]{fontenc}
\usepackage[latin9]{inputenc}
\usepackage[a4paper]{geometry}
\usepackage{babel}
\usepackage{dcolumn}
\usepackage{bm} 
\usepackage{amsmath,amssymb,amsfonts}
\usepackage{verbatim}
\usepackage{graphicx,wrapfig,lipsum}
\usepackage{hyperref}
\usepackage{caption}
\usepackage{bbold}
\usepackage{float} 
\usepackage[usenames,dvipsnames]{xcolor}
\usepackage{epsfig}
\usepackage{epstopdf}
\usepackage{tikz}
\usetikzlibrary{calc,shapes.geometric, arrows}
\usepackage{upgreek}
\usepackage{setspace}
\usepackage{enumitem}
\usepackage{array,multirow,bigdelim}
\usepackage{comment}

\begin{document}
\widetext

\title{On the amplitude expansion of gluon correlators in $\textrm{AdS}_4$}
\author{Humberto Gomez$^{a,b,c}$}
\author{Renann Lipinski Jusinskas$^{b}$}
\author{Arthur Lipstein$^{d}$}
\author{Cristhiam Lopez-Arcos$^{e}$}

\affiliation{$^{a}$Facultad de Ciencias Basicas,  Universidad Santiago de Cali,\\
Calle 5 $N^\circ$  62-00 Barrio Pampalinda, Cali, Valle, Colombia}
\affiliation{$^{b}$Institute of Physics of the Czech Academy of Sciences \& CEICO \\ 
Na Slovance 2, 18221 Prague, Czech Republic}
\affiliation{$^{c}$Physics Department, S\~ao Carlos Federal University,\\
Rodovia Washington Lu\'is, km 235, S\~ao Carlos - SP, Brazil}
\affiliation{$^{d}$Department of Mathematical Sciences, Durham University, Durham, DH1 3LE, UK}
\affiliation{$^{e}$Escuela de Matem\'{a}ticas, Universidad Nacional de Colombia Sede Medell\'{i}n,\\
Carrera 65 $\#$ 59A--110, Medell\'{i}n, Colombia}

\begin{abstract}

We show that tree-level gluon correlators in $\textrm{AdS}_4$  admit a natural expansion in terms of flat-space scattering amplitudes at all multiplicities. In particular, every $n$-point correlator can be decomposed into a sum over energy poles whose residues are flat-space amplitudes. The $n$-point amplitude encodes the flat-space limit while curvature corrections are captured by lower-point amplitudes with merged external data. The merging of external polarizations is recursively defined via an AdS analogue of the Berends-Giele currents, giving rise to all-multiplicity formulae which we verify against Feynman diagram computations up to five points. Crucially, our approach works at the level of full correlators rather than individual diagrams, providing an elegant and transparent alternative to conventional approaches for computing correlators in anti-de Sitter space. 

\end{abstract}

\maketitle

\section{Introduction}

The study of correlators in anti-de Sitter (AdS) space occupies a central role in modern theoretical physics. It  underpins the AdS/CFT correspondence and provides a controlled setting in which to explore the structure of quantum gravity and gauge theories at strong coupling. In recent years, momentum-space methods have proven particularly powerful, revealing that boundary correlators in AdS (and their de Sitter (dS) counterparts, related through a Wick rotation of the radial direction \cite{Maldacena:2002vr,Bzowski:2013sza}) share deep structural similarities with flat-space scattering amplitudes. A prominent example is the flat space limit, in which flat space amplitudes can be obtained by Fourier transforming correlators along boundary directions and taking the residue around the sum of the magnitudes of the boundary momenta \cite{Raju:2012zr}. This observation naturally raises the question: how much flat-space data is embedded in the full correlator?

In this paper we answer this question for Yang-Mills theory in $\textrm{AdS}_4$. We show that every tree-level gluon correlator admits a decomposition into a sum over energy poles, where the residue of each pole is a flat-space scattering amplitude: either the full $n$-point amplitude, or lower-point amplitudes with a precise merging of the external data. The lower-point amplitudes are obtained from the original one by concatenating subsets of external boundary momenta and polarizations. Their energy poles are then obtained by summing over the magnitudes of the merged boundary momenta. 

The first key insight underlying our construction is that the merged polarizations can be interpreted as genuine flat-space polarizations in the Lorenz gauge. They are Lorentz-covariant, associated to solutions of the free equations of motion, and enjoy a residual gauge symmetry. The second key insight is that the recursion satisfied by the multi-particle currents  takes precisely the form of the Berends--Giele (BG) recursion in flat space \cite{Berends:1987me} in terms of merged external data. Together, these two observations allow us to establish all-multiplicity formulae for gluon correlators in $\textrm{AdS}_4$ in terms of flat-space amplitude data alone.

An important feature of our approach is that it operates at the level of full correlators, rather than individual Feynman diagrams. The amplitude expansion structure only becomes manifest when all contributions to a given correlator are assembled. Our formulation makes this structure explicit from the outset, providing a more efficient and transparent route to holographic observables. Our results can also be adapted to cosmological wavefunction coefficients, complementing and extending the cosmological bootstrap program \cite{Baumann:2022jpr}. 

The paper is organized as follows. We begin with a detailed treatment of $\mathrm{Tr}(\phi^3)$ theory in flat space with a boundary, which serves as a transparent testing ground for the key ideas. In particular, we show that the boundary correlators of this theory admit an amplitude expansion, where merged polarizations are now Lorentz scalars, and match this with explicit Feynman diagram calculations up to six points. Next, we describe our main results for Yang--Mills theory in $\textrm{AdS}_4$, including an all-multiplicity formula which we match with Feynman diagram calculations up to five points. We then present some final remarks and future directions to be investigated. In the appendix we describe our underlying framework in terms of classical multi-particle solutions.

\paragraph{Notation:}

We will work in the Poincare patch of AdS$_4$ with metric 
\begin{equation}
ds^{2}=z^{-2} \eta_{mn}dx^{m}dx^{n}=z^{-2}( \eta_{\mu\nu}dx^{\mu}dx^{\nu}+dz^{2}),
\label{adsmetric}
\end{equation}
where $z \geq 0$ describes the radial direction, and $\eta_{\mu \nu}$ is the boundary metric with signature $(-++)$. In practice it will be convenient to perform a Weyl transformation to half of flat space and use the four-dimensional metric $\eta_{m n}$. Indeed, the action of Yang-Mills theory in AdS$_4$ (modulo gauge-fixing terms) is the same as that in flat space due to classical scale invariance.

It is convenient to Fourier transform correlators to momentum space along the boundary. Each external leg (i.e. each single-particle state) will then be assigned a four-momentum $k_m = \{k_\mu, \mathrm{i} k\}$, with $k = \sqrt{\eta^{\mu \nu} k_{\mu} k_{\nu}}$. More generally, we denote multi-particle labels (words) by capital letters, e.g. $P=P_1\ldots P_i$, where the labels $P_j$ denote sub-words. Such labels appear naturally when defining amplitudes with merged external data. Single particle states will be denoted by lower case one-letter words. 

We denote the scalar product between two boundary vectors by $a \cdot b = \eta^{\mu \nu} a_\mu b_\mu$. Boundary momentum is additive in the multi-particle labels, $k^{\mu}_{P}= k^{\mu}_{P_1}+\ldots +k_{P_i}^{\mu}$. For the radial direction, we introduce the energy variables 
\begin{equation}
    E_{P_1,\ldots,P_i}=k_{P_1}+\ldots + k_{P_i}.
\end{equation}
with $k_P=\sqrt{k_P \cdot k_P}$ \footnote{The interpretation in terms of energy is more transparent in de Sitter space, where the bulk direction is timelike and the boundary is Euclidean}. Whereas momentum is conserved along the boundary directions, this is not the case for energy. Finally, we define Lorentz-invariant Mandelstam variables in terms of (merged) momenta
\begin{equation}\label{eq:GeneralSij}
    S_{P_{1},\ldots,P_{i}} \equiv 
  k^2_{P_1\ldots P_i} - E_{P_1,\ldots,P_i}^2,
\end{equation}
which will be recurrent in our construction. 

In this paper, we will consider correlators of matrix-valued scalar fields and Yang-Mills fields. They can be stripped of their matrix structure and expressed in terms of cyclically symmetric color-ordered correlators, which we focus on in practice (see for example \cite{Dixon:1996wi} for a more detailed discussion of color-ordering).     

\section{$\rm{Tr}(\phi^3)$ Theory}

As a toy model, let us consider $\textrm{Tr }(\phi^3)$ theory in half of flat space, where the field $\phi$ is an $N\times N$ matrix with one index in the fundamental and another index in the anti-fundamental representation of $\mathrm{SU}(N)$. The action is given by
\begin{equation}\label{eq:phi3-action}
    S=\int d^{3}xdz\,\mathrm{Tr}\left(\frac{1}{2}\partial_{\mu}\phi\partial^{\mu}\phi+\frac{1}{2}\partial_{z}\phi\partial_{z}\phi-\frac{\lambda}{3}\phi^{3}\right),
\end{equation}
where $z\geq 0$. For simplicity, we will set $\lambda=1$ in the following. Using a Weyl transformation, this can be mapped to a conformally coupled scalar theory in $\textrm{AdS}_4$ with a $z$-dependent cubic interaction (see e.g. \cite{Arkani-Hamed:2017fdk}).

Let us now consider tree-level boundary correlators with Dirichlet boundary conditions. These can be derived from Feynman diagrams of conformally coupled scalars in AdS$_4$ (see for example \cite{Albayrak:2020isk} for more details). As mentioned above, we will work with color-ordered correlators, in which we strip the matrix form of the external polarizations and leave behind only scalar placeholders, which are usually set to $1$. For example, at three-points we have
\begin{equation}
    \mathcal{C}(1, 2, 3)=\frac{1}{E_{1,2,3}} \phi_1 \phi_2 \phi_3,
    \label{phic3pt}
\end{equation}
where the coefficient of the energy pole is simply the flat space three-point amplitude $A(1,2,3)=\phi_1 \phi_2 \phi_3$.

The four-point correlator is given by 
\begin{equation}
\mathcal{C}(1,2,3,4) = \frac{\phi_{1}\phi_{2}\phi_{3}\phi_{4}}{E_{1,2,3,4} E_\textrm{L} E_\textrm{R}}+(\{1,2,3\}\to \{2,3,1\}),\label{eq:4ptphi3-usual}
\end{equation}
where $E_\textrm{L}=E_{1,2,34}$ and $E_\textrm{R}=E_{12,3,4}$ are known as partial energy singularities. In the flat space limit, $E_{1,2,3,4}\to 0$, the product $E_\textrm{L} E_\textrm{R}$ coincides with the Mandelstam variable $S_{1,2}$ and we obtain
\begin{equation}
\lim_{E_{1,2,3,4}\to 0} \mathcal{C}(1,2,3,4) = \frac{1}{E_{1,2,3,4}} A (1,2,3,4),
\label{4ptflat}
\end{equation}
where the coefficient of the energy pole is the $\textrm{Tr }(\phi^3)$ amplitude in flat space:
\begin{equation}
    A (1,2,3,4)=\left(\frac{1}{S_{1,2}} + \frac{1}{S_{2,3}}\right) \phi_{1}\phi_{2}\phi_{3}\phi_{4}.
\end{equation}

Interestingly, if we subtract \eqref{4ptflat} from \eqref{eq:4ptphi3-usual} the remaining part can be expressed in a similar form. To see this, first note that
\begin{equation}\label{eq:identityE}
\frac{1}{E_{1,2,3,4} \, E_\textrm{L}\,  E_\textrm{R}} = \frac{1}{S_{1,2}}\left(\frac{1}{E_{1,2,3,4}} -\frac{1}{E_{12,3,4}} \right).
\end{equation}
Using this identity, we can recast the four-point partial correlator as
\begin{multline}
\mathcal{C}(1,2,3,4)=\frac{A (1,2,3,4)}{E_{1,2,3,4}} + \frac{A (12,3,4)}{E_{12,3,4}} +\frac{A (1,23,4)}{E_{1,23,4}}. \label{eq:4ptphi3-flat}
\end{multline}
Note that the second and third terms are sub-leading in the flat space limit. The coefficients of these sub-leading energy poles are given by flat space three-point amplitudes with merged external momenta and polarizations, labeled by ${ij}$. In particular a merged momentum has boundary components $k^\mu_{ij}=k^\mu_{i}+k^\mu_{j}$ and energy $k_{ij}= \sqrt{k_{ij}\cdot k_{ij}}$, such that the four-momentum is null. The merged scalar polarization is expressed as the Lorentz-invariant object 
\begin{equation}
    \Phi_1(ij) = - \frac{1}{S_{i,j}}  \phi_i \phi_j, \label{eq:phi3-merdeg2pt}
\end{equation}
which is simply the cubic vertex containing the polarizations of legs $i$ and $j$ with the flat space propagator attached to it. In \cite{Baumann:2020dch}, the residues of partial energy singularities of certain four-point wavefunction coefficients in $\textrm{dS}_4$  were shown to factorize into three-point amplitudes times linear combinations of lower-point wavefunction coefficients. We note here that they can be naturally interpreted as flat space polarizations. 

The structure of equations \eqref{eq:4ptphi3-flat} and \eqref{eq:phi3-merdeg2pt} extends to arbitrary multiplicity. Its generalization employs multi-particle coefficients $\Phi_{a}(P_{1},\ldots,P_{a})$ analogous to BG currents, but adapted to flat space with a boundary. They are recursively obtained via
\begin{multline}
    S_{P_1,\ldots,P_a}  \Phi_{a}(P_{1},\ldots,P_{a}) \\ = \sum_{b=1}^{a-1} \Phi_{b}(P_{1},\ldots,P_{b})\Phi_{a-b}(P_{b+1},\ldots,P_{a}).
    \label{eq:Phi-recursion}
\end{multline}
Note that four-dimensional Lorentz invariance is still manifest. The main difference from the usual construction in flat space is that we now have the merged polarizations built into the recursion, denoted by $\Phi_1(P)$. For one-letter words, they correspond to the usual single-particle polarizations, $\Phi_1(i)=\phi_i$. Otherwise, they are given by
\begin{equation}
\Phi_{1}(P)	=	-\sum_{a=2}^{|P|}\sum_{P=P_{1}\cdots P_{a}} \Phi_{a}(P_{1},\ldots,P_{a}), \label{eq:Phi3-merged}
\end{equation}
where $|P|$ denotes the length of the word $P$, and the sum $\sum_{P = P_{1}\ldots P_{a}}$ denotes a sum over all possible deconcatenations of the word $P$ into $a$ non-empty subwords $P_b$, with $b=1,\ldots,a$. This expression is tightly connected to Dirichlet boundary conditions, and the details of its derivation are presented in  Appendix \ref{BGdetails}.

Using the multi-particle coefficients of \eqref{eq:Phi-recursion}, we can then define flat-space $n$-point amplitudes \`a la BG,
\begin{multline}\label{eq:phi3-flatamp}
A(P_{1},\ldots,P_{n-1},n) \\=  \phi_n S_{P_1,\ldots,P_{n-1}}\Phi_{n-1}(P_{1},\ldots,P_{n-1}).
\end{multline}
which, in turn, become the coefficients of the energy-pole expansion of $n$-point partial correlators, expressed as
\begin{equation}
  \mathcal{C}(1,2,\ldots,n)=\sum_{a=2}^{n-1} \, \sum_{1\cdots n-1=P_{1}\cdots P_{i}}\frac{A (P_1,\ldots,P_a,n)}{E_{P_{1},\ldots,P_{a},n}}.
\label{eq:phi3npt}
\end{equation}
Note that the only terms in this expansion which are not Lorentz-invariant are the energy poles. In this form, the flat space limit ($E_{1,2,\ldots,n} \to 0$) manifestly yields the $n$-point amplitude, $A_{n} (1,2,\ldots,n)$, and the sub-leading contributions are cast as a sum over lower-point energy poles whose residues correspond to amplitudes with merged external data. In more conventional approaches this structure is partially hidden and corrections to flat space limit are determined using bootstrap techniques \cite{Arkani-Hamed:2018kmz,Baumann:2020dch,Goodhew:2020hob,Jazayeri:2021fvk,Melville:2021lst}. As usual in the BG construction, the external leg $n$ is special in the sense that it is the one that is attached to the BG current. As a result, we express the amplitudes solely in terms of Mandelstam invariants which do not depend explicitly on leg $n$. We have checked that \eqref{eq:phi3npt} agrees with standard Feynman diagram calculations up to six-points \footnote{A Mathematica notebook with these comparisons is attached to the arXiv submission}. Note that the amplitude expansion has spurious singularities which arise when external momenta become collinear (sometimes known as folded singularities). At four points the cancellation of such singularities can be seen from \eqref{eq:identityE}, but at higher points it arises from a very non-trivial cancellation among all the terms in the sum in \eqref{eq:phi3npt}.  


\section{Yang-Mills Theory}

Now let us consider Yang-Mills theory in $\textrm{AdS}_4$:
\begin{equation}
    S_{\textrm{YM}}=\int d^{3}xdz \,\sqrt{-g}g^{mn}g^{pq}F_{mn}F_{pq},
\end{equation}
where $F_{mn}=\partial_m A_n - \partial_n A_m - \mathrm{i}[A_m,A_n]$, and $A_m$ is the gauge field. Using the metric in the Poincar\'e patch, $g_{mn}=z^{-2} \eta_{mn}$,  we see that all factors of $z$ cancel out and the action reduces to that in half of flat space:
\begin{equation}\label{eq:YM-action}
    {S}_{\textrm{YM}}=\int d^{3}xdz \, \eta^{mn}\eta^{pq}F_{mn}F_{pq}.
\end{equation}
This reflects the classical conformal invariance of Yang-Mills theory in four dimensions. We will use this form of the action to compute tree-level color-ordered correlators.

Gluon correlators in AdS are usually computed via Feynman diagrams in the axial gauge ($A_z=0$), with Dirichlet boundary conditions \cite{Liu:1998ty,Raju:2011mp,Maldacena:2011nz}. The three-point correlator is given by
\begin{multline}
    \mathcal{C}(1,2,3) = \frac{1}{E_{1,2,3}}[(k_{1}-k_{2})\cdot\varepsilon_{3}](\varepsilon_{1}\cdot\varepsilon_{2})\\+\textrm{cyclic}\{1,2,3\}, \label{eq:YM3pt}
\end{multline}
where $\varepsilon^\mu_p$ denote the boundary components of the single-particle polarizations, which satisfy $k_p \cdot \varepsilon_p=0$. Moreover the four-point partial correlator can be expressed as (see for example \cite{Albayrak:2018tam})
\begin{multline}
    \mathcal{C}(1,2,3,4)\\= \frac{1}{E_{1,2,3,4} E_{\textrm{L}}E_{\textrm{R}}} V_{\mu}(1,2)\left(\eta^{\mu\nu}+\frac{k_{12}^{\mu}k_{34}^{\nu}}{k_{12}^{2}}\right)V_{\nu}(3,4)\\+\frac{1}{E_{1,2,3,4}}V_{\textrm{C}}-(\{1,2,3\}\leftrightarrow\{2,3,1\}), \label{eq:YM4pt}
\end{multline}
where
\begin{equation}
    V_{\mu}(i,j)=(\varepsilon_i \cdot \varepsilon_j)k_{i\mu}+2(k_j \cdot \varepsilon_i) \varepsilon_{j\mu}- (i\leftrightarrow j),
\end{equation}
encodes the three-point vertex. Note that the first line on the right hand side contains the transverse part of the gluon exchange. The longitudinal part appears as an effective contribution to the contact interaction, given by
\begin{multline}
    V_{\textrm{C}}=(\varepsilon_{1}\cdot\varepsilon_{3})(\varepsilon_{2}\cdot\varepsilon_{4})-(\varepsilon_{1}\cdot\varepsilon_{4})(\varepsilon_{2}\cdot\varepsilon_{3})\\-(\varepsilon_{1}\cdot\varepsilon_{2})(\varepsilon_{3}\cdot\varepsilon_{4})\frac{(k_{1}-k_{2})(k_{3}-k_{4})}{k_{12}^{2}}. \label{eq:YM-contact}
\end{multline}

In direct analogy with the $\rm{Tr}(\phi^3)$ theory, these correlators admit an expansion in terms of flat space amplitudes. Remarkably, when these amplitudes are obtained using BG recursion in the Lorenz gauge ($\partial^{m}A_{m}=0$), the resulting correlators agree with Feynman diagram calculations in axial gauge provided we fix the residual gauge transformations \footnote{Note that we are referring to Lorenz gauge in half of flat space. In AdS this corresponds to a non-covariant gauge-choice.}. This is evident in the three-point case of equation \eqref{eq:YM3pt}, where the coefficient of the energy pole $E_{1,2,3}$ is the known three-point YM amplitude when $\varepsilon^z_p=0$.

As for the four-point correlator, it can be recast as
\begin{multline}
\mathcal{C}(1,2,3,4)=\frac{1}{4 E_{12,3,4}}A({12},3,4)+\frac{1}{4 E_{1,23,4}}A(1,{23},4)\\+\frac{1}{4 E_{1,2,3,4}}A(1,2,3,4) + \textrm{cyclic}\{1,2,3,4\}, \label{eq:YM-4ptcorrelator}
\end{multline}
where $A$ denotes flat space partial amplitudes computed via the traditional BG currents \cite{Berends:1987me}. In particular $A(1,2,3,4)$ is the four-point amplitude, and $A({12},3,4)$ and $A(1,{23},4)$ are three-point amplitudes with merged external momenta and polarizations $\mathcal{A}_m^{(1)}(ij)$, given by
\begin{subequations}\label{eq:2particle-merged}
\begin{align}
    \mathcal{A}^{(1)}_{\mu}(ij) &= - \frac{1}{S_{i,j}} V_{\mu}(i,j),\\
    \mathcal{A}_{z}^{(1)}(ij) &	=  \frac{\mathrm{i}}{k_{ij}}  [k_{ij}\cdot \mathcal{A}^{(1)}(ij)].
\end{align}
\end{subequations}
The boundary components of the merged polarizations are obtained by combining three-point vertices containing two polarizations with the inverse Mandelstam to which they are attached. The radial component then follows from the Lorenz gauge condition. Hence, they satisfy the expected physical conditions ($k^m_{ij} k_{ijm} = k^m_{ij} \mathcal{A}_m^{(1)}(ij)=0$). We will provide a more detailed comparison of \eqref{eq:YM-4ptcorrelator} to \eqref{eq:YM4pt} after introducing some more notation. 

Let us now state how the amplitude expansion in \eqref{eq:YM-4ptcorrelator} extends to arbitrary multiplicity. This can be phrased in terms of generalized BG currents $\mathcal{A}_{m}^{(a)}(P_{1},\ldots,P_{a})$ which are recursively defined via
\begin{widetext}
    \begin{multline}\label{eq:YM-BG}
S_{P_{1},\ldots,P_{a}} \mathcal{A}_{m}^{(a)}(P_{1},\ldots,P_{a})=\sum_{b=1}^{a-1}\mathcal{A}_{n}^{(b)}(P_{1},\ldots,P_{b})\mathcal{A}_{p}^{(a-b)}(P_{b+1},\ldots,P_{a})V_{m}^{np}(P_{1},\ldots,P_{b}|P_{b+1},\ldots,P_{a})\\+\sum_{b=1}^{a-2}\sum_{c=b+1}^{a-1}\mathcal{A}_{n}^{(b)}(P_{1},\ldots,P_{b})\mathcal{A}_{p}^{(c-b)}(P_{b+1},\ldots,P_{c})\mathcal{A}_{q}^{(a-c)}(P_{c+1},\ldots,P_{a})(\delta_{m}^{q}\eta^{np}-\delta_{m}^{n}\eta^{pq}-2\delta_{m}^{p}\eta^{nq}).
    \end{multline}
\end{widetext}
The first line comes from the cubic interaction, with
\begin{multline}
    V_{m}^{np}(P_{1},\ldots,P_{b}|P_{b+1},\ldots,P_{a})=2k_{P_{b+1},\ldots,P_{a}}^{n}\delta_{m}^{p}\\-2k_{P_{1},\ldots,P_{b}}^{p}\delta_{m}^{n}+\eta^{np}(k_{P_{1},\ldots,P_{b}m}-k_{P_{b+1},\ldots,P_{a}m}),
\end{multline}
where $k^m_{P_{1},\ldots,P_{a}}=k^m_{P_{1}}+\ldots+k^m_{P_{a}}$, while the second line encodes the quartic interactions.

In spite of this inflated appearance, equation \eqref{eq:YM-BG} is just a generalization of the usual flat space BG recursion in which the words of the arguments ($P_{1},\ldots,P_{i}$)  play the role of external-particle labels. This is compatible with the interpretation of $\mathcal{A}_{m}^{(1)}$ as merged polarizations. For one-letter words, they coincide with the single-particle polarizations, i.e. $\mathcal{A}_{\mu}^{(1)}(i)=\varepsilon_{i\mu}$ and $\mathcal{A}_{z}^{(1)}(i)=0$. More generally, they take a similar form to equation \eqref{eq:Phi3-merged}. In the boundary directions, we have
\begin{subequations}
\label{eq:YM-mergedpol}
\begin{equation}
\mathcal{A}_{\mu}^{(1)}(P)	=-\sum_{a=2}^{|P|}\sum_{P=P_{1}\ldots P_{a}}\mathcal{A}_{\mu}^{(a)}(P_{1},\ldots,P_{a}),\label{eq:YM-mergedpol-mu}
\end{equation}
while the component in the radial direction is determined by imposing the Lorenz gauge, such that
\begin{equation}
\mathcal{A}_{z}^{(1)}(P)	=  \frac{\mathrm{i}}{k_P}  [ k_P \cdot  \mathcal{A}^{(1)}(P)].
\end{equation}     
\end{subequations}
These merged polarizations are associated with solutions to the free equations of motion (see Appendix \ref{BGdetails}), and they even enjoy a residual gauge transformation of the form $\delta \mathcal{A}_{m}^{(1)}(P) = k_{Pm} \Lambda_P$. 

In analogy with the flat space construction, we then define the following ordered amplitudes:
\begin{multline}
    A({P}_{1},\ldots,{P}_{a},n) \equiv \varepsilon^m_{n} [S_{P_1,\ldots,P_a}  \mathcal{A}^{(a)}_m(P_1,\ldots,P_a)].
\end{multline}
Note that they have exactly the same form as scattering amplitudes obtained via BG recursion in flat space without a boundary when written in terms of merged polarizations. In this case, however, we do not have momentum conservation along the radial direction so we do not impose energy conservation. In terms of these building blocks, tree-level gluon correlators are then given to arbitrary multiplicity via
\begin{multline}
\mathcal{C}(1,\ldots,n) = \frac{1}{n}\sum_{a=2}^{n-1}\sum_{1\ldots n-1=P_{1}\ldots P_{a}}\frac{A({P}_{1},\ldots,{P}_{a},n)}{E_{P_{1},\ldots,P_{a},n}} \\ + \textrm{cyclic}\{1,2,\ldots,n\}, \label{eq:ym-Npcorrelator}
\end{multline}
in which we explicitly add the cyclic averaging of the external polarizations. Note that such a sum is not needed in flat space without a boundary due to  momentum conservation. This was not necessary in the $\rm{Tr}(\phi^3)$ theory because the interactions do not contain derivatives. As in the $\rm{Tr}(\phi^3)$ case, leg $n$ is special and we express the amplitudes solely in terms of Mandelstam invariants which do not depend explicitly on leg $n$. The correlator \eqref{eq:ym-Npcorrelator} is written in Lorenz gauge. Under residual gauge transformations, we generate boundary contact terms, which are proportional to delta functions in position space \cite{Maldacena:2011nz,Armstrong:2020woi}.

Now we have all the ingredients to show that the four-point expressions in \eqref{eq:YM4pt} and \eqref{eq:YM-4ptcorrelator} agree. To illustrate how this works, let us begin with the first term in \eqref{eq:YM4pt}, which describes a gluon exchange, and apply the identity in \eqref{eq:identityE}:
\begin{multline}
    \frac{1}{E_{1,2,3,4}E_{\textrm{L}}E_{\textrm{R}}}V_{\mu}(1,2)V^{\mu}(3,4)\\
    =\left(\frac{1}{E_{1,2,3,4}}-\frac{1}{E_{12,3,4}}\right)\frac{1}{S_{1,2}}V_{\mu}(1,2)V^{\mu}(3,4),\\
       =\frac{1}{E_{1,2,3,4}}\mathcal{A}_{\nu}^{(2)}(1,2)\varepsilon_{3\rho}\varepsilon_{4}^{\mu}V_{\mu}^{\nu\rho}(1,2|3)\\+\frac{1}{E_{12,3,4}}\mathcal{A}_{\nu}^{(1)}(12)\varepsilon_{3\rho}\varepsilon_{4}^{\mu}V_{\mu}^{\nu\rho}(12|3).
\end{multline}
We are then left with two terms which we can easily recognize from the amplitude expansion in \eqref{eq:YM-4ptcorrelator}. In particular the first term comes from the amplitude $A(1,2,3,4)$ computed via BG recursion (times the total energy pole), while the second term simply corresponds to the three-point amplitude $A(12,3,4)$ times its corresponding energy pole, where $\mathcal{A}_{\nu}^{(1)}(12)$ is a merged polarisation. Similarly, the first line of $V_{\textrm{C}}$ in \eqref{eq:YM-contact} comes from the current $\mathcal{A}_{m}^{(3)}(1,2,3)$ which contributes to $A(1,2,3,4)$ in \eqref{eq:YM-4ptcorrelator}. 

We have also checked the general formula in \eqref{eq:ym-Npcorrelator} against Feynman diagrams at five points. In this case, it can be compactly expressed as 
\begin{multline}
    \mathcal{C}(1,\ldots,5)=\frac{1}{5}\bigg\{\frac{A(1,2,3,4,5)}{E_{1,2,3,4,5}}+\frac{A({12},3,4,5)}{E_{12,3,4,5}}\\+\frac{A(1,{23},4,5)}{E_{1,23,4,5}}+\frac{A(1,2,{34},5)}{E_{1,2,34,5}}+\frac{A({123},4,5)}{E_{123,4,5}}\\+\frac{A({12},{34},5)}{E_{12,34,5}}+\frac{A(1,{234},5)}{E_{1,234,5}}\bigg\}+\textrm{cyclic}\{1,\ldots,5\}.
\end{multline}
For $\varepsilon^{z}_{i} = 0$ and the merged polarizations in \eqref{eq:YM-mergedpol}, this non-trivially agrees with diagrammatic computations in axial gauge \cite{Albayrak:2018tam,Albayrak:2019asr}.

\section{Discussion}

In this work, we have established that the full tree-level correlator of gluons in $\textrm{AdS}_4$ is organized by flat-space amplitude data at all multiplicities. The energy-pole expansion we developed makes this structure explicit and transparent, bypassing the complexity of individual diagram evaluations. The lower-point amplitudes appearing in this expansion are defined in terms of merged external momenta and polarizations, where the merged polarizations satisfy the expected properties of genuine (single-particle) external states and are defined via recursion relations. This gives rise to all-multiplicity formulae which we verified against Feynman diagram computations in the $\textrm{Tr}(\phi^3)$ theory and in Yang-Mills in the axial gauge.

The gluon scattering amplitudes which appear in these expansions are computed using Berends-Giele recursion in the Lorenz gauge. Remarkably, the resulting correlators match Feynman diagram calculations of pure Yang-Mills theory in axial gauge. Since energy is not conserved, amplitudes obtained using other methods or gauge choices will yield correlators which differ by boundary terms arising from total derivative terms in the bulk Lagrangian. A similar observation extends to $\textrm{Tr}(\phi^3)$ theory, given that there are different ways of rewriting the flat space amplitude using momentum conservation. It would be interesting to study the relation between the choice of representatives in the amplitude expansion and total derivative terms in the bulk action more systematically.

We believe the most immediate next step in this framework is to formulate our gluonic correlators in terms of spinor-helicity variables \cite{Maldacena:2011nz}. The merged polarizations generally comprise a linear combination of helicities. As a result, the analogue of the Parke--Taylor formula \cite{Parke:1986gb} in our setting would involve a sum over lower-point amplitudes with different helicity configurations weighted by the corresponding lower-point energy poles. It would be natural to express them in terms of four-dimensional (momentum) twistors associated with flat space (taking into account the lack of energy conservation) \cite{Hodges:2009hk,Adamo:2011pv}, twistors adapted to AdS$_4$ \cite{Bittleston:2024rqe,Baumann:2024ttn}, or the orthogonal Grassmannian \cite{Arundine:2026fbr}.

Beyond gauge theory, the extension of the amplitude expansion to correlators of non-conformal theories in AdS$_4$, most notably Einstein gravity, may require replacing the flat-space amplitudes by certain differential operators acting on the corresponding energy poles. Such a representation could provide useful new perspectives on color/kinematics duality and the double copy in AdS$_4$ \cite{Armstrong:2020woi,Albayrak:2020fyp,Armstrong:2023phb}. We have already observed these structures in the scalar toy models related to the self-dual sectors of Yang-Mills and Einstein gravity in AdS$_4$ (see \cite{Lipstein:2023pih,Chowdhury:2024dcy}), and this will be reported in an upcoming paper \cite{upcoming}. There we will also discuss non-trivial relations between the amplitude-based representation for gluon correlators introduced here and a geometric formulation of certain scattering amplitudes in flat space known as the ABHY associahedron \cite{Arkani-Hamed:2017mur}. One may also ask whether this extends recent geometric formulations such as \cite{Arkani-Hamed:2017fdk,Gomez:2021qfd,Arkani-Hamed:2024jbp} to spinning correlators, and whether one could explore them beyond tree-level. Taken together, these directions suggest that flat-space scattering amplitudes are not merely a limiting case of AdS correlators, but a fundamental organizing principle underlying quantum field theory in curved spacetime.

\begin{acknowledgments}
We thank Guilherme Pimentel and Evgeny Skvortsov for useful discussions. H.G is partially supported by FAPESP, grant 2026/00820-8.  R.L.J. is supported by the GA\v{C}R grant 25-16244S from the Czech Science Foundation. A.L. is supported by an STFC Consolidated Grant ST/T000708/1.
\end{acknowledgments}

\appendix

\section{Multi-particle solutions and correlators} \label{BGdetails}

Here we introduce the classical multi-particle solutions associated to the actions \eqref{eq:phi3-action} and \eqref{eq:YM-action}, which are an extension of the perturbiner framework \cite{Rosly:1996vr,Rosly:1997ap,Mafra:2015gia} in flat space (see also \cite{Armstrong:2022mfr} for its implementation in AdS). We present their recursive definition taking into account boundary conditions, and explain how they lead to the correlator equations \eqref{eq:phi3npt} and \eqref{eq:ym-Npcorrelator}.

The classical equation of motion associated to \eqref{eq:phi3-action} is simply (for $\lambda=1$)
\begin{equation}
    (\eta^{\mu \nu} \partial_\mu \partial_\nu +\partial^2_z) \phi = - \phi^2,
\end{equation}
and we will consider the multi-particle ansatz
\begin{equation}
    \phi(z) =\sum_{P}\Phi_{P}(z)e^{ik_{P}\cdot x}.
\end{equation}
The solution is recursively determined by substituting the ansatz into the equation of motion. We then obtain a differential equation of the form
\begin{equation}\label{eq:phi3-multi}
    (k^2_P -\partial^2_z) \Phi_P = \sum_{P=QR} \Phi_Q \Phi_R.
\end{equation}
Notice that color-ordering is implicit in the deconcatenation structure of the sum. In order to invert the operator on the left hand side, we need the appropriate Green's function, $G(z,y)$, satisfying
\begin{equation}
    (k^2 -\partial^2_z) G_k(z,y) = \delta(z-y).
\end{equation}
This is where the boundary condition comes in. We then impose  Dirichlet's condition, and obtain
\begin{equation}
    G_{k}(z,y)=\frac{1}{2k}\left(e^{-k|z-y|}-e^{-k(z+y)}\right),
\end{equation}
with $G(0,y)=G(z,0)=0$.

Instead of solving the differential equation \eqref{eq:phi3-multi}, we can simply expand $\Phi_P(z)$ in the eigenfunctions of the kinetic operator. It can be written as
\begin{equation}
    \Phi_{P}(z)=\sum_{a=1}^{|P|}\sum_{P=P_{1}\ldots P_{a}}\Phi_{a}(P_{1},\ldots,P_{i})e^{-E_{P_{1},\ldots,P_{a}}z}.
\end{equation}
The Dirichlet boundary condition is then implemented by requiring $\Phi_{P}(0)=0$ for $|P|>1$. It leads to
\begin{equation}
    \Phi_{1}(P)+\sum_{a=2}^{|P|}\sum_{P=P_{1}\ldots P_{a}}\Phi_{a}(P_{1},\ldots,P_{a})=0,
\end{equation}which we then solve for $\Phi_{1}(P)$. This is the origin of equation \eqref{eq:Phi3-merged}. Note, in particular, that $\Phi_{1}(P)$ does not appear  on the left hand side of equation \eqref{eq:phi3-multi}. It parametrizes a free solution to the equation of motion, which leads to its interpretation as a merged polarization. 

From the equation of motion, we derive the recursion for the coefficients $\Phi_{i}(P_{1},\ldots,P_{i})$, displayed in equation \eqref{eq:Phi-recursion}. It matches the usual recursion in flat space without a boundary, with labels $P_b$ being associated to merged polarizations. Analogously, the definition of the $(n+1)$-point correlator is a direct generalization of the one for flat space amplitudes, and can be cast as
\begin{multline}\label{eq:phi3-original-correlator}
    \mathcal{C}(1,2,\ldots,n+1)\propto\int d^{3}x\int_{0}^{\infty}dz\left(\phi_{n+1}e^{\mathrm{i}k_{n+1}\cdot x}e^{-k_{n+1}z}\right)\\ \times \left(k_{N}^{2}-\partial_{z}^{2}\right)\Phi_{N}(z)e^{\mathrm{i}k_{N}\cdot x},
\end{multline}
where $N=1\ldots n$. The overall proportionality constant can be fixed, for instance, with a direct comparison to the three-point correlator computed using Feynman rules. In flat space, the integrals would simply yield a momentum conserving delta function, which now happens only in the boundary directions. The integration over the radial coordinate $z$ yields the energy poles, finally leading to
\begin{multline}
  \mathcal{C}(1,2,\ldots,n+1)=\sum_{a=2}^{n} \, \sum_{1\cdots n=P_{1}\cdots P_{a}}\frac{1} {E_{P_{1},\ldots,P_{a},n+1}}\\ \times \phi_{n+1 } S_{P_1,\ldots,P_a} \Phi_a (P_1,\ldots,P_a),
\end{multline}
which matches equation \eqref{eq:phi3npt} after the identification \eqref{eq:phi3-flatamp}.

For Yang--Mills, which is described by the action in \eqref{eq:YM-action}, the procedure is very similar to the scalar case, so we will just discuss the relevant differences. The equation of motion in Lorenz gauge ($\partial^m A_m=0$) can be cast as
\begin{multline}
   (\eta^{\mu \nu} \partial_\mu \partial_\nu +\partial^2_z) A_{m}=2\mathrm{i}[A^{n},\partial_{n}A_{m}]\\-\mathrm{i}[A^{n},\partial_{m}A_{n}]+[[A_{m},A_{n}],A^{n}],
\end{multline}
and we will work with the color-stripped ansatz
\begin{equation}\label{eq:YM-multi-ansatz}
    A_{m}(x,z)=\sum_{P}\mathcal{A}_{Pm}(z)e^{\mathrm{i}k_{P\mu}x^{\mu}}T^{a_{P}},
\end{equation}
where $T^{a_{P}} = T^{a_{p_1}} \ldots T^{a_{p_n}}$ for the word $P=p_1 \ldots p_n$, denoting a product of gauge group generators.

Just like in the scalar theory, we introduce the eigenfunction expansion
\begin{equation}
    \mathcal{A}_{mP}(z)=\sum_{a=1}^{|P|}\sum_{P=P_{1}\ldots P_{a}}\mathcal{A}_{m}^{(a)}(P_{1},\ldots,P_{a})e^{-E_{P_{1},\ldots,P_{a}}z}.
\end{equation}
By inserting this ansatz into the equation of motion, we obtain the recursion for the currents displayed in equation \eqref{eq:YM-BG}. We also note that $\mathcal{A}_{m}^{(1)}(P)$ is the coefficient of a free solution ($k_{Pm} k^{m}_{P}=0$).

For Dirichlet boundary conditions, we would expect $\mathcal{A}^m_{P}(0)=0$ for $|P|>1$. However, this is not compatible with the Lorenz gauge. Using equation \eqref{eq:YM-mergedpol}, we see that $\mathcal{A}^\mu_{P}(0)=0$. However, for the radial direction we have that $\partial_z \mathcal{A}^z_{P}(z)$ vanishes at the boundary, which is identified with a Neumann boundary condition. It is remarkable that the result of BG recursion in Lorenz gauge matches axial gauge Feynman diagrams with Dirichlet boundary conditions. Similar observations were made in \cite{Skvortsov:2026gtq}. Note that the solution \eqref{eq:YM-mergedpol} is not unique, as $\mathcal{A}_{m}^{(1)}(P)$ parametrizes a free solution to the equation of motion, enjoying the same residual gauge invariance of single-particle polarizations, $\delta \mathcal{A}_{m}^{(1)}(P)= k_{mP} \Lambda_P$. 


The $(n+1)$-point correlator is a generalization of the scalar case, but with an important caveat. Since momentum is not conserved along the radial direction, and Yang--Mills theory has derivative interactions, a naive application of \eqref{eq:phi3-original-correlator} breaks the cyclicity of the correlator. This is the reason why we implement a cyclic average in the definition of the YM correlator in \eqref{eq:ym-Npcorrelator}.

\bibliography{references}

\end{document}